\pdfoutput=1
\documentclass[9pt,twoside]{rmaa-rho_light}
\RMxAAtemplatetype{\RMxAA} 

\newcommand{\beq}{\begin{equation}}
\newcommand{\eeq}{\end{equation}}

\newcommand{\hii}{H~{\sc II}~}
\newcommand{\kms}{\mbox{km s$^{-1}$}}

\newcommand{\Mo}{\mbox{M$_{\odot}$}}
\newcommand{\Moy}{\mbox{M$_{\odot}$ yr$^{-1}$}}

\vol{62}
\pages{1-5}
\thisyear{2026}
\doi{\href{https://www.astroscu.unam.mx/rmaa/RMxAA..XX-X}{https://www.astroscu.unam.mx/rmaa/RMxAA..XX-X}}

\title{Analytical Solutions for the Dynamics of Planetary Nebulae with and without Common Envelope Evolution}

\author[]{G.~ García-Segura \orcidlink{0000-0002-9899-2292}}

\affil[]{Instituto de Astronomía, Universidad Nacional Autónoma de México, Apdo. Postal 106, 22800 Ensenada, B.C., Mexico}

\leadauthor{García-Segura}
\smalltitle{Analytical Solutions for Planetary Nebulae}

\corres{Guillermo García-Segura}
\email{ggs@astro.unam.mx}

\received{\today}
\accepted{\today}

\license{Texto de la licencia aquí}

\setbool{rho-abstract}{true} 
\setbool{rho-resumen}{true} 

\begin{abstract}
We present new analytical solutions for the dynamics of planetary nebulae.
These equations consider the temporal variation of the mechanical luminosity as well as the common envelope evolution scenario. By comparing a database of nebulae with these solutions,
a large portion of planetary nebulae can be better explained by the common envelope evolution scenario, especially the fast and  slow ones. Single AGB stellar models can only reproduce nebulae with expansion velocities between 20 and 30 $\kms$.

\end{abstract}


\begin{resumen}

Presentamos unas soluciones analíticas nuevas para la dinámica de nebulosas planetarias. 
Estas ecuaciones contemplan la variación temporal de la luminosidad mecánica así como el escenario
de evolución de envoltura común. Contraponiendo una base de datos de nebulosas con estas ecuaciones,
una gran porción de nebulosas planetarias pueden ser mejor explicadas con el 
escenario de evolución de envoltura común, especialmente las rápidas y las lentas.
Los modelos estelares de AGB individuales sólo pueden reproducir nebulosas con velocidades de expansión entre 20 y 30 $\kms$.

\end{resumen}

\begin{document}

\maketitle
\pagestyle{fancy}\thispagestyle{firststyle}


\section{INTRODUCTION}

The kinematical studies of nebulae in general has been of interest throughout astronomy, from \hii regions, interstellar bubbles, supernova remnants, and planetary nebulae (PNs).
Interstellar bubbles of massive stars were studied with a continuous injection of wind in the works of \citet{Avedisova72}, \citet{Dyson77} and \citet{Weaver77}, as well as PNs in the works of  \citet{Kwok78} and \citet{Volk85} for low mass stars. More analytical solutions for all kind of cases can be found in the work of \citet{Ostriker88}. A similar study to that of PNs was applied to 
Wolf-Rayet nebulae in \citet{GGS95}.

In the specific case of PNs, different observational works
have studied the relationship between their sizes and expansion velocities, such as
those in \citet{Chu84}, \citet{Sabba84}, \citet{Weinberger89}, \citet{Bianchi92}, and
more recently in \citet{Icker21}.

In the majority of previous studies in the literature, it is assumed that the expansion rate of PNs is constant, as dictated by the solutions of \citet{Kwok78}. However, as we will see below,
the mechanical luminosity of the fast winds of PNs is not
constant \citep{Villaver2002II}. Therefore, new equations are needed to understand their behavior. This is the first objective in this article.

On the other hand, PNs not only evolve from an isolated asymptotic giant branch (AGB) star, but the Common Envelope Evolution (CEE) scenario can also occur in the case of binaries \citet{LivioSoker88}.
This implies that most of the AGB's envelope is ejected suddenly and abruptly
in a few years, compared to tens or hundreds of thousands of years in the case of isolated AGBs. As can be seen in \citet{Villaver2002I}, the largest mass-loss rates in the AGB occur
in the later phases, starting at 50\% of the AGB lifetime for high-mass stars (5 \Mo), or 75\% of the lifetime for low-mass stars (1 \Mo).
Therefore, depending on when the CEE occurs, the mass of the ejected envelope
can approach 100\% in early CEE events in the AGB phase. This, obviously,
depends on the separation of the binary.
Furthermore, when the envelope is ejected, not all of it is completely ejected in the sense
of being gravitationally unbound, but up to 75\% of the envelope
can remain gravitationally bound \citep{Ricker2012}. That is, it remains in orbit
around the binary system. In the previous study, most of the mass is ejected at an angle of approximately 45 degrees around the equator, but in other calculations with different physics, not all the gas is ejected towards the equator; in fact, the result is almost spherical, as is the case in \citet{Ivanova18}. There is still no consensus on the distribution of the ejected gas, how much is completely ejected from the system, whether there is any uniformity, or if it depends on each specific case. See \citet{DeMarco25} for a review on the subject. Although the CEE scenario is very complex, we need to develop a simple model that we can use as a first approximation. This is the second objective of this article. 

In this article we will treat PNs as pressurized bubbles by central stellar winds. Other scenarios as those that included jets \citep{SokerLivio94} are not taken into account, 
although jets could also be treated as a wind in certain specific cases \citep{Akashi2025}. 
The effects of the
\hii region is also not taken into account (see numerical calculations in \citet{GGS25}).

This article is structured as follows: the analytical solutions for the
Single AGB progenitor are described in \S~2. The solutions for a CEE progenitor are 
described in \S~3. 
Comparisons with observed PNs  are presented in \S~4.   Finally, we discuss the results in \S~5 and provide the main conclusions in the last section.

\section{SINGLE AGB PROGENITOR}

In this scenario it is assumed that the AGB star evolves directly to the PN phase. The
mass loss at the AGB phase produces an external medium in which the density drops as
$\rho \sim r^{-2}$. Then, the PN central star (PNCS) lunches a fast wind. 
The mechanical luminosity of the fast wind is
\beq L_{\rm w} \,\,=  \,\, \frac{1}{2}\,\, \dot{M} \,\, v_{\infty}
^2  \,\,, \eeq
where $\dot{M}$ is the mass-loss rate of the PNCS, and $v_{\infty}$ 
is the final wind velocity at the arrival of the reverse shock. 
$L_{\rm w}$ exceeds the mechanical luminosity of the AGB wind  by
a factor of 100-1000.  To allow
a continuous variation in the transition from AGB to PN, we can model the time
dependent mechanical luminosity as
\beq L_{\rm w}  \,\, = \,\,\, L_0 \,\,\left(\frac{t}{t_0}\right)^{\delta}
\,\,\, = \,\,\, {\cal L}_0 \,\,\, t^{\delta}\,\,. \eeq

The AGB wind velocity is two orders of magnitude smaller than the PNCS wind
velocity, so it can be neglected in the dynamical equations. 
In order to get self-similar solutions, we neglect the external pressure,
giving us the dynamical equations
\beq {\frac{d}{d t}} \left [ M_{\rm s}(t) \dot R_{\rm s}(t)
\right ] = 4 \pi R_{\rm s}^2 P, \eeq
\beq {\frac{d E} {d t}} = L_{\rm w} - 4 \pi R_{\rm s}^2 P \dot R_{\rm s} \,\,, \eeq
where $P = E / 2 \pi R_{\rm s}^3$ and $E$ are the thermal pressure and the thermal energy of the hot, shocked, PN wind region, and
\beq M_{\rm s}(t) =  \frac{\dot{M}_{\rm AGB}} {v_{\rm AGB}}\, R(t) , \eeq 
is the swept-up shell mass. This system of equations can be solved by
assuming the swept-up shell to be thin, so that the contact surface
between the hot interior and the cold shell is located at nearly the 
same radius $R_{\rm s}$ as the outer shock. The solution is
\beq E \, = \, \left( \frac{ 3 {\cal L}_0}{5 \delta + 9}\right)\,\,\,t^{\delta + 1} , \eeq

\beq R_{\rm s} \, = \, 3 \left( \frac{2 \, v_{\rm AGB}}{\dot{M}_{\rm AGB}} \right)^
{1/3} \left( \frac{1}{2 \delta^2 + 3 \delta + 9 }\right)^{1/3} 
\left( \frac{ {\cal L}_0}{5 \delta + 9 } \right)^
{1/3}  t^{(\delta + 3) / 3} ,\eeq

\beq \dot R_{\rm s}  =  \frac{\delta + 3}{3} \left( \frac{2 \, v_{\rm AGB}}{\dot{M}_{\rm AGB}} \right)^
{1/3} \left( \frac{1}{2 \delta^2 + 3 \delta + 9 }\right)^{1/3} 
\left( \frac{ {\cal L}_0}{5 \delta + 9 } \right)^
{1/3}  t^{\delta / 3} , \eeq

where $\gamma$ = 5/3  is used as the adiabatic index of the hot shocked PN gas, and
$t$ is the time measured from the beginning of the PN phase. 
Equation (6) gives the thermal energy that remains in the hot, shocked
wind region, equation~(7) is the equation of motion for the swept-up shell, and equation~(8) gives the expansion velocity of the PN. Here, the velocity of the AGB wind $v_{\rm AGB}$ must
be added to the solution, since the expansion velocity starts from this value at $t=0$. Equation~(8) shows that a mechanical luminosity increasing in 
time ($\delta > 0$) gives accelerating bubbles, while a luminosity decreasing
in time ($\delta < 0$) gives decelerating bubbles.  The case of $\delta = 0 $
gives constant expansion velocity in this case.

The main difference with the previous study by \citet{Volk85} (in their section 2.3) is that 
we have chosen to fix the 
AGB wind as a $\rho \sim r^{-2}$ because, during the last thermal pulses in the stellar evolution calculations of \citet{Vas94}, the mass-loss rates and wind velocities of the AGBs are fairly constant in time \citep{Villaver2002I}, so this approximation works well in this case. This choice allows to arrive at an explicit analytical expression for the radius and expansion velocity. 
This would be the case of $ \beta=-2$  in \citet{Volk85}.

\section{CEE PROGENITOR}

In this scenario, the AGB envelope is assumed to be ejected abruptly. Part of the envelope
will be ejected from the binary system, and the other big portion of that will remain gravitationally bound. 
Thus, the mass of the shell $M_{\rm s}$ will be this bound mass, and it is located at a distance very close to the binary system. We assume that this shell has a constant mass ($M_{\rm s} = {\rm constant} $) throughout  its evolution, and it is assumed that the shell will not interact with the ejected portion of the envelope that was not gravitationally bound exceeding the escape velocity. With this approach, we solve again equations~(3)  and ~(4), and the solution is

\beq E \, = \, \left( \frac{  {\cal L}_0}{2 \delta + 4}\right) \,\,\,t^{\delta + 1} , \eeq

\beq R_{\rm s} \, = \, \left( \frac{8}{2 \delta^3 + 12 \delta^2  + 22 \delta + 12 }\right)^{1/2} 
\left( \frac{ {\cal L}_0}{M_{\rm s}} \right)^
{1/2}  t^{(\delta + 3) / 2} , \eeq

\beq \dot R_{\rm s}  =  \left( \frac{\delta + 3}{2} \right)  \left( \frac{8}{2 \delta^3 + 12 \delta^2  + 22 \delta + 12 }\right)^{1/2} \left( \frac{ {\cal L}_0}{M_{\rm s}} \right)^
{1/2}  t^{(\delta + 1 ) / 2}. \eeq

Equation~(9) gives the thermal energy of the hot, shocked
wind region, equation~(10) is the equation of motion of the shell, and equation~(11) gives the expansion velocity of the post-CEE PN. 
Equation~(11) shows that a mechanical luminosity constant in time ($\delta = 0$)
gives accelerating PNs. Note that only values for  $\delta > -1$ are allowed in these equations.

\begin{figure}[ht]
\centering
\includegraphics[width=0.99\columnwidth]{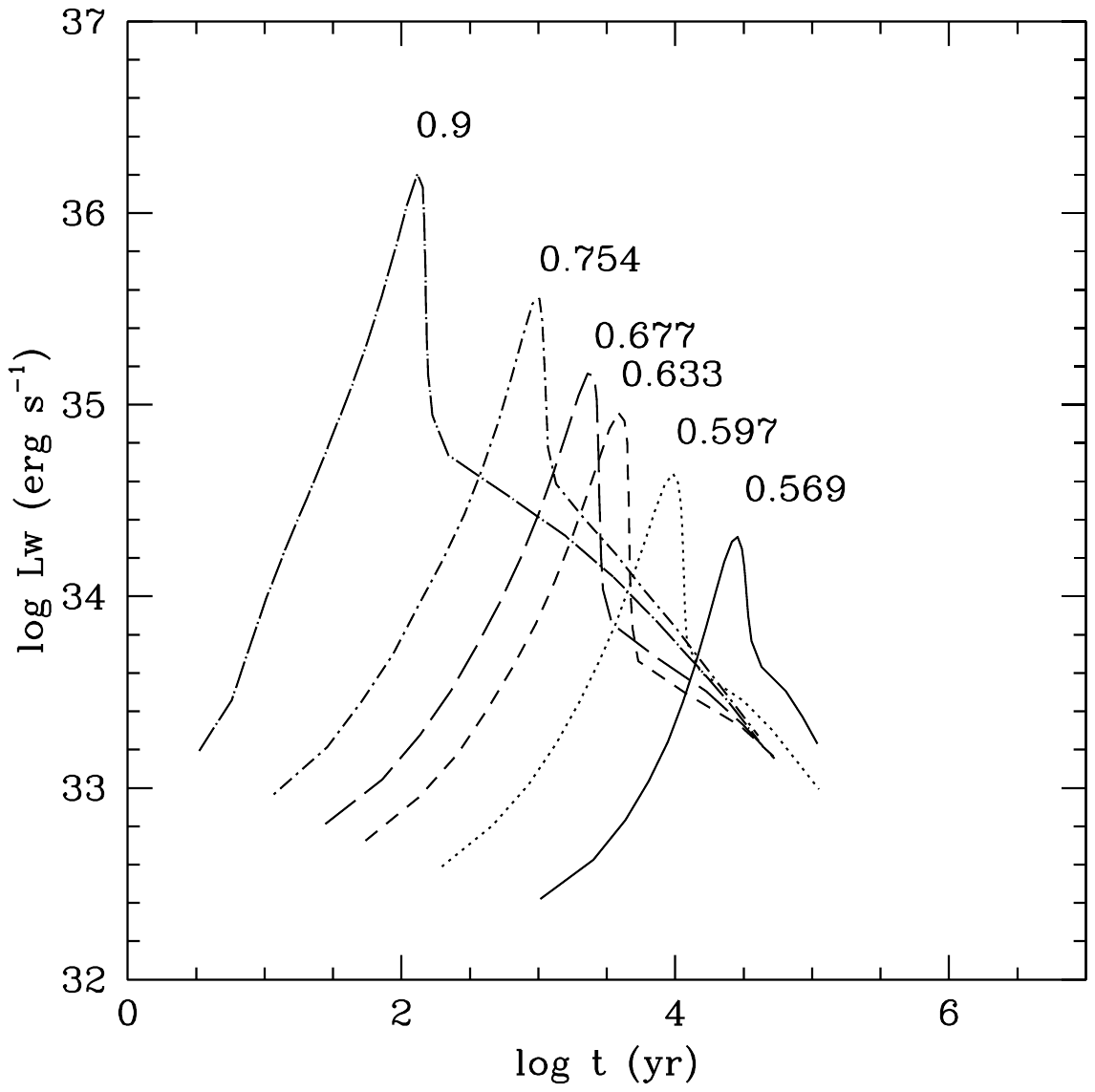}
\caption{Mechanical Luminosity as a function of time from models by \cite{Vas94}. The numbers are
the mass of the central star.}
\label{fig:figure1}
\end{figure}

\begin{figure}[ht]
\centering
\includegraphics[width=0.99\columnwidth]{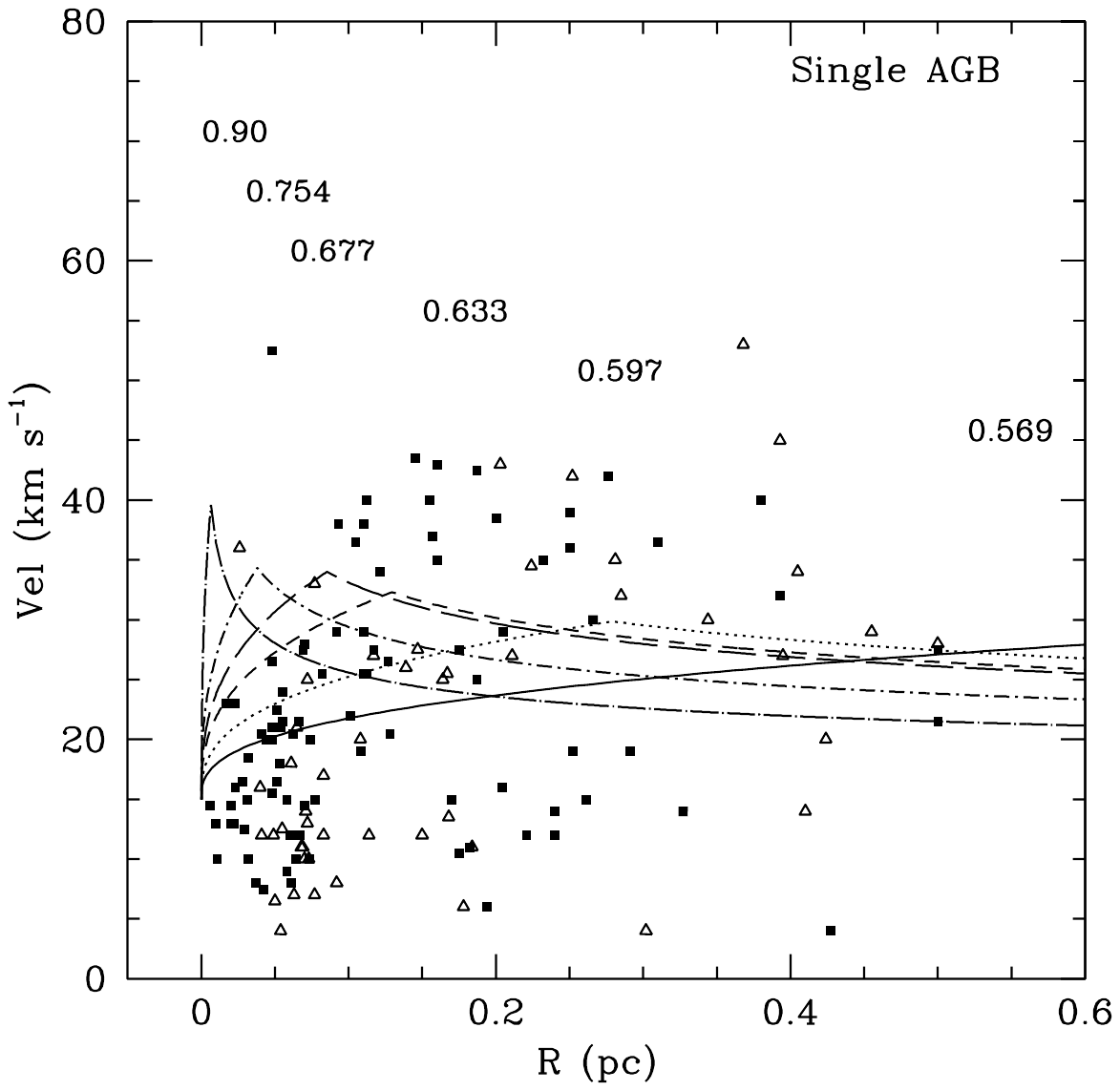}
\caption{Results for the single AGB scenario. The solid squares correspond to the data in \citet{Sabba84}, while the
open triangles to the data by \citet{Icker21}.}
\label{fig:figure2}
\end{figure}

\begin{figure}[ht]
\centering
\includegraphics[width=0.99\columnwidth]{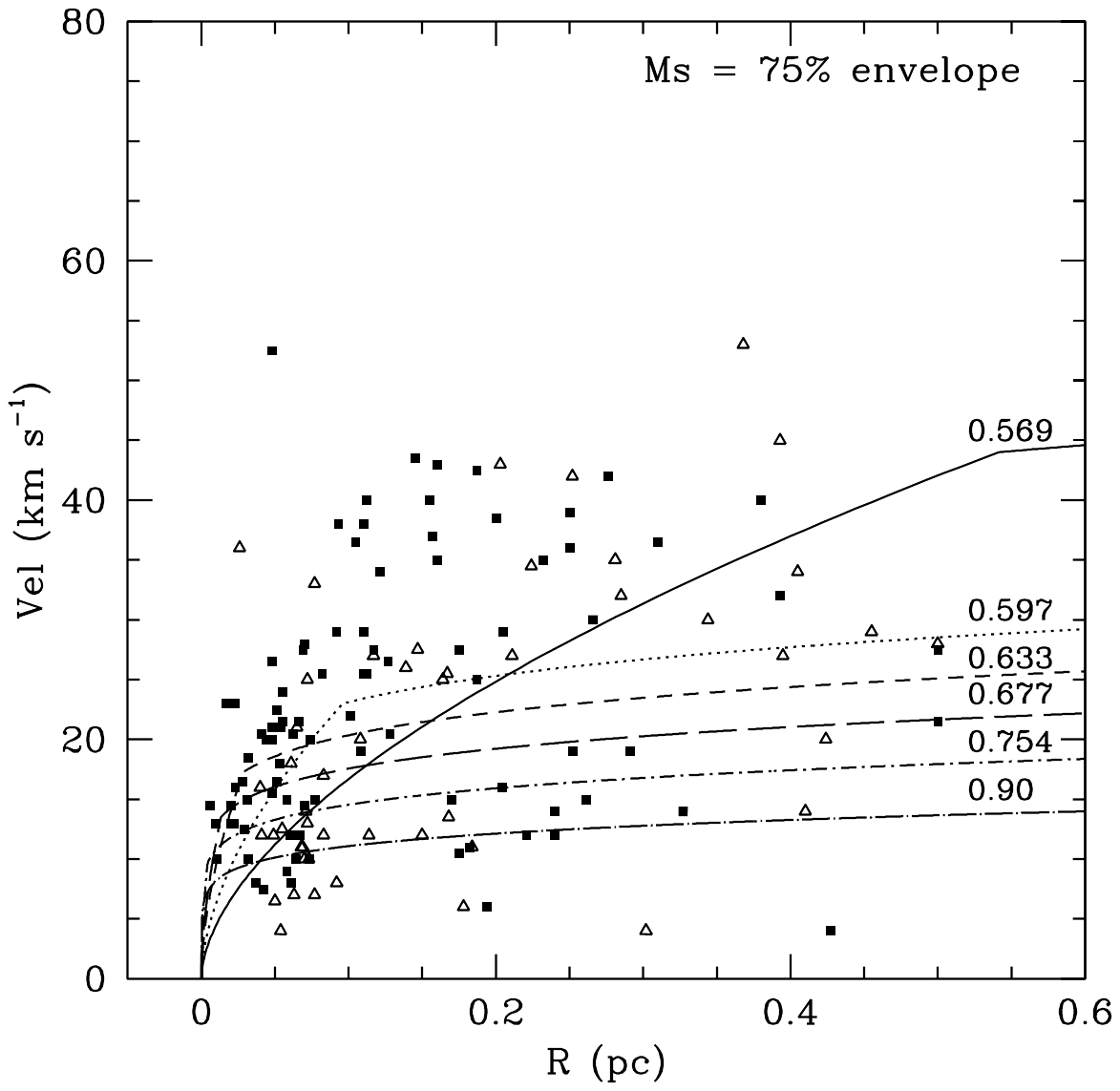}
\caption{Same as Figure 2 for the case of a CEE scenario with 75 \% of the envelope mass in the shell.}
\label{fig:figure3}
\end{figure}

\begin{figure}[ht]
\centering
\includegraphics[width=0.99\columnwidth]{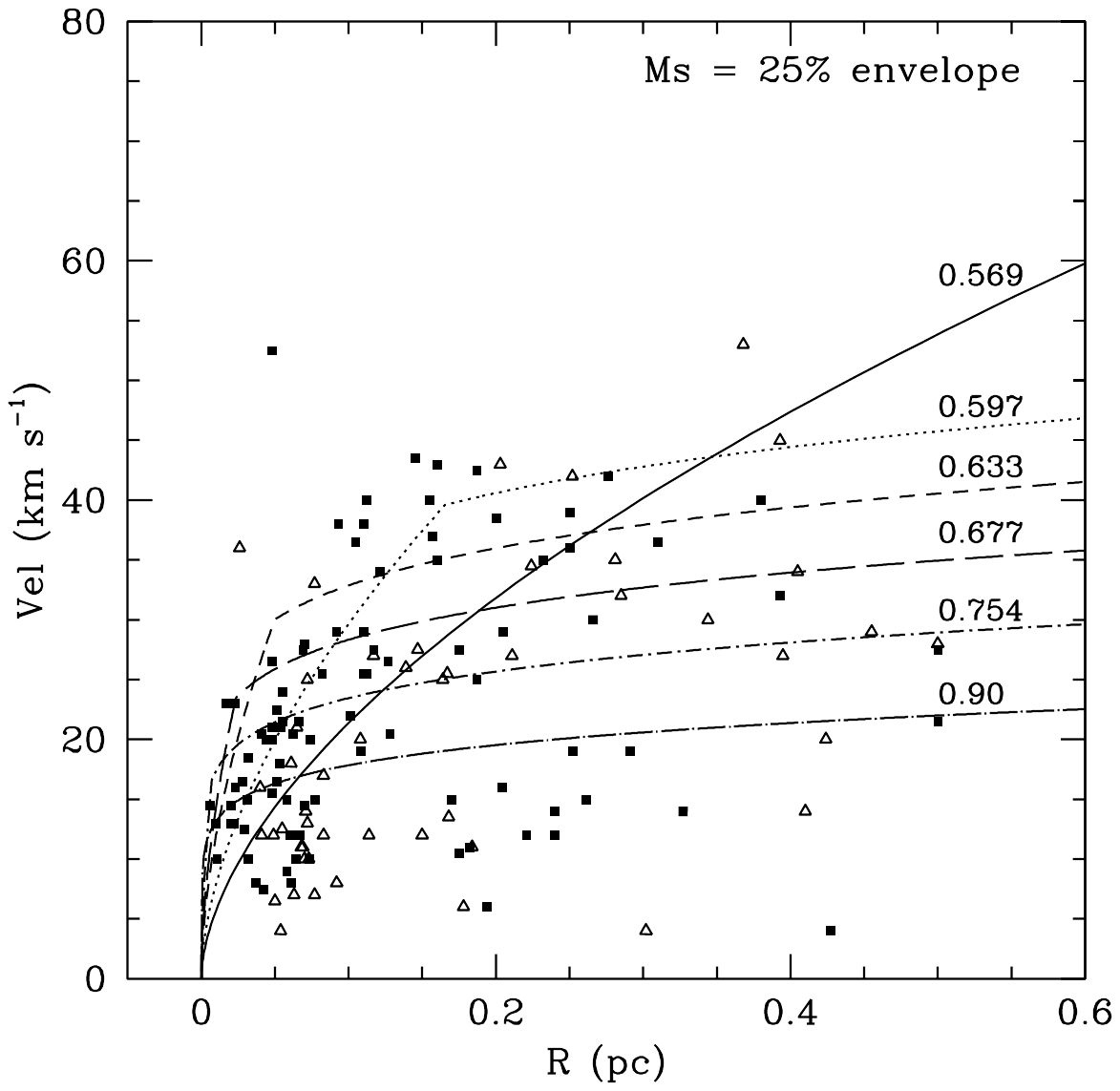}
\caption{Same as Figure 2 for the case of a CEE scenario with 25 \% of the envelope mass in the shell.}
\label{fig:figure4}
\end{figure}

\begin{figure}[ht]
\centering
\includegraphics[width=0.99\columnwidth]{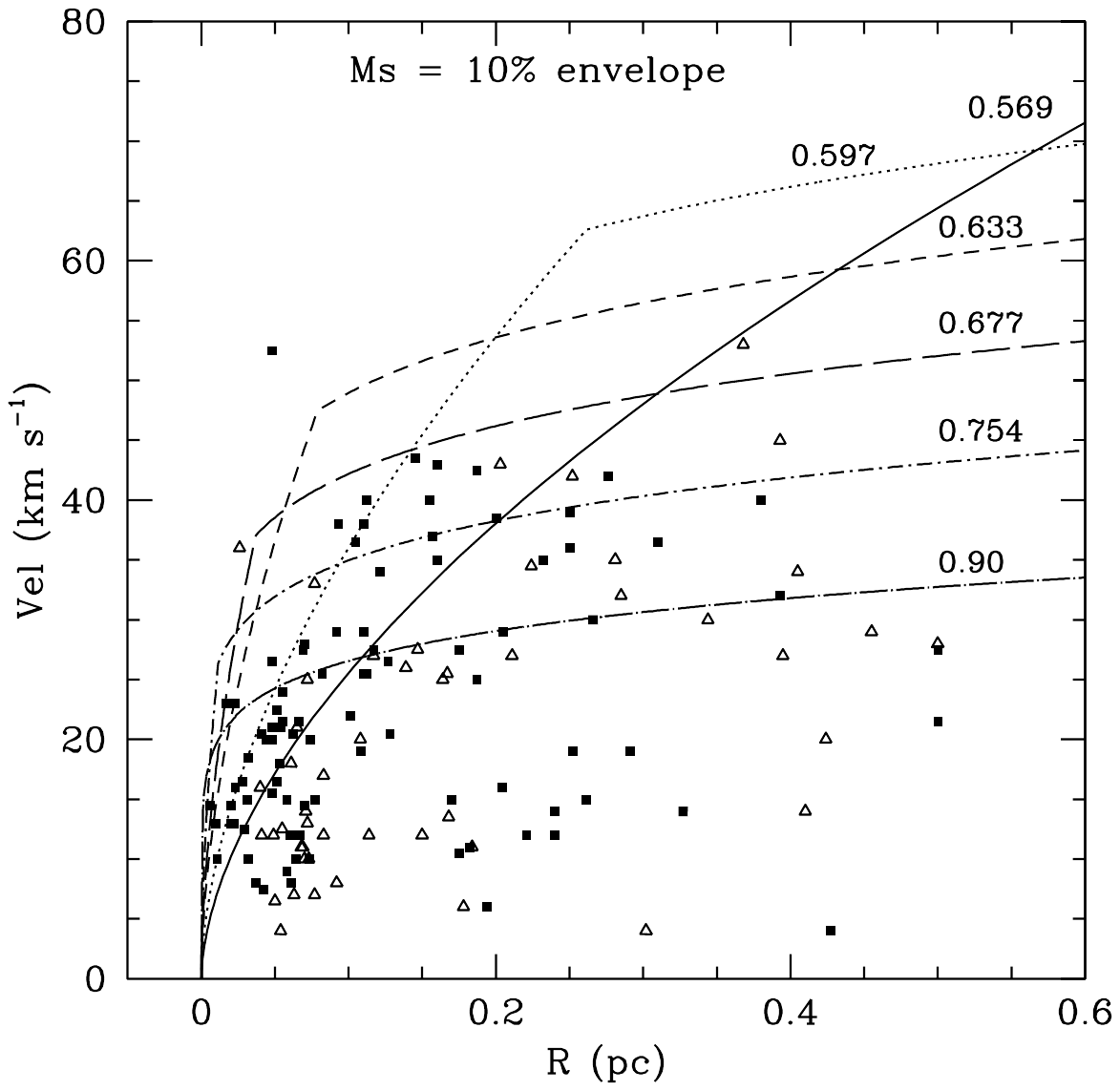}
\caption{Same as Figure 2 for the case of a CEE scenario with 10 \% of the envelope mass in the shell.}
\label{fig:figure5}
\end{figure}

\section{COMPARISON WITH OBSERVED PLANETARY NEBULAE}

To compare with observations of PNs, we first need to manage how to incorporate the mechanical luminosity 
into the equations.
Figure 1 shows the mechanical luminosities of six  models \citep{Vas94} for different stellar masses. 
We can
notice that they have initially a steep increase, and once the highest point is reached, it descends along the cooling track. For simplicity, since we treat the mechanical luminosity as a power law (equation 2), we will fit two straight lines for the mecanical luminosity, one for the ascent and another one for the descent. For the ascent, we can fit a straight line with a slope $\delta = 1.7 $, and for the descent, we fit a slightly smoother drop with a slope $\delta = -0.7$. The
short, abrupt descent at the beginning of the cooling track is not considered for simplicity.

Table 1 gives the inputs used for our comparison, either for the single AGB precursor or for the 
CEE one. Equations ~(7) and ~(8) and equations ~(10) and  ~(11)  are applied respectively for each scenario
with $\delta = 1.7 $ up to the time $t_{\rm CT}$ where the central star begins the excursion to the cooling track. After that point is reached, we use the power laws 
$ R = R_{\rm CT} (t/t_{\rm CT})^{(\delta + 3) / 3} $ and
$ \dot R = \dot R_{\rm CT} (t/t_{\rm CT})^{\delta / 3} $ 
with $\delta = -0.7$ for the single AGB progenitor, or  
$ R = R_{\rm CT} (t/t_{\rm CT})^{(\delta + 3) / 2} $ and
$ \dot R = \dot R_{\rm CT} (t/t_{\rm CT})^{(\delta + 1 ) / 2} $
for the CEE progenitor, where $R_{\rm CT}$ and $\dot R_{\rm CT}$ are the radius and the velocity
at  $t_{\rm CT}$.

Figure 2 shows the case of the single AGB progenitor, 
while Figures 3, 4 and 5 show the
CEE scenarios with 75\%, 25\% and 10\% of the envelope mass in the swept-up shells.
The solid squares correspond to the data in \citet{Sabba84}, while the
open triangles to the data by \citet{Icker21}.
Note that the maximum shell mass of 75\%  correspond to the result obtained in the
calculation by \citet{Ricker2012}, where 75\% of the ejected mass remains gravitationally bound,
and the rest 25\% of the mass is totally ejected from the binary system.

\begin{table*}[hb]
\caption{Parameters for each stellar mass model (CS = Central Star, CT = Cooling Track, E = Envelope). \\
The AGB mass-loss rate and AGB wind velocity are the average of the last thermal pulse.  ${\cal L}_0$ is for $\delta = 1.7$ \,.} 
\label{tab:anysymbols}
\centering
\begin{tabular}{lcccccccc}
\toprule
$M_{\rm ZAMS}$ & $M_{\rm CS}$  & $\dot{M}_{\rm AGB}$ & $ v_{\rm AGB}$ & 
${\cal L}_0$  & $t_{\rm CT}$ & $ 75\% \,\, M_{\rm E}$ & $25\% \,\, M_{\rm 
E}$ &  $ 10\% \,\, M_{\rm E}$   \\
$\Mo$ & $\Mo$ & $\Moy$ & $\kms$ & ${\rm erg \, s^{-1} s^{-1.7} }$ & yr & $\Mo$ & $\Mo$ & $\Mo$ \\
\midrule
1  & 0.569 & $5.31 \times 10^{-6}$& 15 & $1.42 \times 10^{14}$ & 28,600 & 0.32 & 0.1 & 0.043  \\
1.5 & 0.597 & $7.88 \times 10^{-6}$ & 15 & $1.49 \times 10^{15}$ & 9,624 & 0.677 & 0.225 & 0.09 \\
2  & 0.633 & $1.09 \times 10^{-5} $ & 15 & $1.56 \times 10^{16}$ & 3,866 & 1.025 & 0.34 & 0.136  \\
2.5 & 0.677 & $ 1.11 \times 10^{-5} $ & 15 & $5.05 \times 10^{16}$ & 2,297 & 1.367 & 0.455 & 0.182 \\
3.5 & 0.754 & $ 1.85 \times 10^{-5} $ & 15 & $3.57 \times 10^{17}$ & 1,017 & 2.059 & 0.686 & 0.274  \\
5   & 0.900  & $ 3.00 \times 10^{-5} $ & 15 &  $3.92 \times 10^{19}$ & 129 & 3.075 & 1.025 & 0.41  \\
\bottomrule
\end{tabular}
\end{table*}

\section{DISCUSSION}

Let's start our discussion with Figure 2. The first thing that stands out is that single
AGB stellar models can only reproduce nebulae with expansion velocities between 20 and 30 $\kms$. This is extremely interesting, since in the past they have always been compared with constant-velocity models. Another striking feature is the inversion that occurs with respect to central mass; that is, nebulae with larger stellar mass begin to expand much faster than those with lower mass, given their bigger mechanical luminosity. However, when they begin to decelerate on the cooling track, the more massive models start to expand more slowly than the less massive ones. This is because the more massive models have a very short lifetime in their winds
because of the faster evolution. 

The model with a shell of 75\% of the envelope mass in Figure 3 accurately reproduces the early kinematics of many nebulae, especially those with low velocities, in fact, velocities lower than the AGB wind. These nebulae with velocities lower than the AGB wind cannot be explained by the single AGB scenario. On the other hand, the 75\% model fails to reproduce the fastest nebulae.
The 25\% model in Figure 4 reproduces a large number of nebulae, although it also fails to reproduce the very fast and very slow ones. Only the 10\% model in Figure 5 can reproduce the very fast nebulae.

This behavior can be understood in three ways. First, that not all ejections in the CEE events are the same, and that in some cases more mass remains in the system than in others; that is, the efficiency of the ejection can vary considerably for different
nebulae.
Second, this behavior could indicate also that many bipolar nebulae originating from a CEE scenario might exhibit this characteristic; that is, if viewed pole-on, we can observe faster velocities. Within this model, this would imply that the amount of mass deposited in the polar regions in the CEE event is small, which is why low-mass models, such as the 10\% model, can reproduce these velocities. On the other hand, bipolar nebulae viewed
face-on show much slower kinematics, as we observe the expansion of the material at the equator. This makes sense in the current model, since a high concentration of ejected gas in the equatorial region is consistent with a high-mass model, such as the 75\% model. For example, the highest velocity point as a solid square  corresponds to the nebula NGC 2392, which is a clear example of a pole-on nebula \citep{Tere2012}, which also exhibit jets \citep{Martin2021} .
Third. It could also be understood within the evolutionary framework, that is, depending on the separation of the binary, if the separation is small, the CEE can occur very early in the AGB, so the envelope is almost intact, and the mass of the shell would be large, while in the case of long separations, the envelope has already lost a lot of mass by the time the CEE occurs, so we would have a smaller mass for the shell.

\section{CONCLUSIONS}

In this article, we have presented a set of simple equations for the dynamics of PNs, where two scenarios for their formation are considered: the isolated AGB star scenario and the CEE scenario. As a first test, the analytical solutions are compared with a sample of observational data, and show that a large portion of planetary nebulae are better explained by the CEE scenario. This first test is very striking, and it can lead to an in-depth study for each object, in order to differentiate from which scenario it could evolve. What's interesting about this test is that it was done in a blind way, without knowing what morphology each object has.
A study combining dynamics with morphology (\citet{SchwarzCorradi1992}, \citet{Manchado1996}) will be  the subject of a future article.

\renewcommand{\refname}{REFERENCES}
\bibliography{ggsbib}

\section{ACKNOWLEDGEMENTS}

We thank the referee for his/her comments that improved the manuscript.
GGS is partially supported by PAPIIT grant IG101223.

\end{document}